\documentclass[11pt,showpacs,preprintnumbers,amsmath,amssymb,prd,nofootinbib,superscriptaddress]{revtex4-2}
\usepackage[utf8]{inputenc}    
\usepackage[T1]{fontenc}       
\usepackage[brazil,english]{babel} 
\usepackage{graphicx}
\usepackage{mathrsfs}
\usepackage{xcolor}
\usepackage{listings}

\newcommand{\bea}{\begin{eqnarray}}
\newcommand{\eea}{\end{eqnarray}}

\begin{document}

\title{On Nonrelativistic Isotropic and Homogeneous Universe}

\author{R.G.G. Amorim}\email[]{ronniamorim@gmail.com }
\affiliation{Department of Physics,  Gama Faculty, University of Brasília,\\
72.444-240, Brasília, DF, Brazil.} 
\affiliation{International Center of Physics, Instituto de F\'isica, Universidade de Bras\'ilia, 70910-900, Bras\'ilia, DF, Brazil.}
\affiliation{Canadian Quantum Research Center,\\ 
204-3002 32 Ave Vernon, BC V1T 2L7,  Canada.} 

\author{A.F. Santos}\email[]{alesandroferreira@fisica.ufmt.br}
\affiliation{Programa de P\'{o}s-Gradua\c{c}\~{a}o em F\'{\i}sica, Instituto de F\'{\i}sica, Universidade Federal de Mato Grosso, Cuiab\'{a}, Brasil}

\author{K.V.S. Araújo}\email[]{kayo.vaz@aluno.unb.br}
\affiliation{Instituto de F\'isica, Universidade de Bras\'ilia, 70910-900, Bras\'ilia, DF, Brazil.} 

\author{S.C. Ulhoa}\email[]{sc.ulhoa@gmail.com}
\affiliation{Instituto de F\'isica, Universidade de Bras\'ilia, 70910-900, Bras\'ilia, DF, Brazil.} 
\affiliation{Canadian Quantum Research Center,\\ 
204-3002 32 Ave Vernon, BC V1T 2L7,  Canada.}

\date{\today}

\begin{abstract}
This article deals with a nonrelativistic cosmological model based on Galilean covariance, formulated within a five-dimensional Galilean manifold. Within this framework, we construct an isotropic and homogeneous metric analogous to the Friedmann--Robertson--Walker metric but without a universal speed limit. Two distinct solutions of the Einstein-like field equations are obtained: (i) a vacuum configuration (\(\lambda=0\)) yielding an exponential--quadratic scale factor, and (ii) a dust-dominated universe (\(\lambda=1\)) described by a non-interacting nonrelativistic fluid. Upon dimensional reduction to 3+1 spacetime through a specific embedding, the model naturally develops anisotropy in the scale factor and density, consistent with the near-zero spatial curvature inferred from Planck data. In the case of vanishing spatial curvature, the framework reproduces Milne’s Newtonian cosmology because this condition leads to a vanishing pressure. This provides an independent nonrelativistic setting for cosmological dynamics within Galilean covariance.
\end{abstract}

\maketitle

\section{Introduction}\label{sec.1}

The theory of General Relativity (GR) stands as the most widely accepted description of gravitation and successfully models cosmology on large scales. However, despite this success, several phenomena at both cosmic and galactic scales remain unexplained within GR. The puzzles of dark energy and dark matter lack intrinsic justification in the classical framework of GR, which has motivated alternative approaches and extended theories of gravity \cite{capozziello2008, ulhoa2012, ulhoa2019}. On the other hand, Newtonian cosmology in the traditional 3+1 formulation has proven highly useful as a simplified dynamical model. Milne in 1934 \cite{milne1934} demonstrated that a homogeneous self-gravitating fluid in Euclidean space could reproduce the expansion dynamics of the universe in a manner similar to Friedmann solutions, although without spacetime curvature. This Newtonian paradigm also introduces the possibility of an absolute reference frame that may provide insight into cosmological effects. The hypothesis that dark energy and dark matter might have contributions arising from reference-frame effects cannot yet be ruled out. Later developments extended Newtonian cosmology to include fluids with non-zero pressure \cite{fabris2013}. These extensions, however, have always been regarded as approximations to GR rather than independent nonrelativistic gravitational frameworks. Such an objective can be achieved through an approach in which Galilean transformations are written in a covariant manner. This approach is known as Galilean covariance.

The usual formulation of Galilean transformations embodies the nonrelativistic symmetry within a standard 3+1 decomposition, lacking a covariant structure and treating time as a separate parameter. However, the Galilean energy relation $E = \frac{p^2}{2m}$ suggests the possibility of a covariant framework, since $p^2 - 2mE = 0$ resembles the norm of a null vector in a five-dimensional space, where the penta-momentum is defined as $P^{\mu} = (\vec{p}, E, m)$ and the metric is adjusted to satisfy $P^2 = \eta_{\mu\nu} P^{\mu} P^{\nu} = 0$. Thus, a five-dimensional manifold endowed with such a metric can reproduce Galilean physics in a covariant manner. In this setup, the three-dimensional momentum integrates with energy in a form analogous to relativistic momentum space, while mass plays the role of the fifth component of the penta-vector within the so-called Galilean manifold. Galilean symmetry can then be expressed through coordinate transformations, and its covariant formulation requires the introduction of a fifth coordinate, whose specific functional choice recovers the usual 3+1 decomposition. Consequently, fields invariant under Galilean symmetry can be described through a covariant structure akin to relativistic physics, albeit with entirely distinct interpretations. This property is particularly advantageous for nonrelativistic systems, as it simplifies the dynamical equations. In other words, Galilean covariance enables a unified treatment of nonrelativistic physics alongside relativistic field theories, while maintaining their independence. For instance, the Schr\"odinger equation emerges as a Klein--Gordon field on the Galilean manifold, and the Navier--Stokes equation can be derived from a least-action principle applied to an interacting scalar field \cite{Santana2000, Santana2003a, Santana2003b, Santana2003c}. The natural extension of Galilean covariance is the formulation of a gravitational theory through the generalization of Riemannian geometry on the Galilean manifold.

Galilean gravity represents a geometric approach to the gravitational interaction within the Galilean manifold, effectively reformulating Newtonian gravitation in a covariant five-dimensional setting \cite{Ulhoa2009}. It is not the only theory pursuing this objective, as illustrated by the Newton--Cartan formulation developed by Duval \textit{et al.}~\cite{duval1985}, yet it remains fully consistent with matter fields described under Galilean covariance. This covariant formalism inherently enables the treatment of nonrelativistic gravitational phenomena without the need for \textit{ad hoc} assumptions, encompassing, for example, the dynamics of stars within galaxies. In this context, it becomes natural to seek a fully nonrelativistic cosmological solution that does not stem from an approximation to General Relativity, as outlined in the earlier part of this section. Since Galilean gravity features five-dimensional equations formally analogous to Einstein’s, the nonrelativistic fluid governed by the Navier--Stokes equations can serve as the matter source on cosmological scales. Consequently, applying the principles of homogeneity and isotropy in nonrelativistic cosmology under Galilean covariance defines a novel approach to Newtonian cosmology. As we shall demonstrate, our results closely reproduce Milne’s original formulation, thereby confirming the consistency of the embedding into the standard 3+1 decomposition adopted here.

The article is organized as follows. In Section~\ref{sec.2}, we present the foundational ideas of Galilean covariance. Section~\ref{sec.3} develops the nonrelativistic cosmological model and examines two distinct cases for the nonrelativistic fluid. Finally, Section~\ref{sec.4} summarizes the main conclusions.

\section{Galilean Covariance} \label{sec.2}

In this section, we present a brief review of the galilean covariance. Our discussion begins with the Galilei symmetry. The ideas presented here are based in references \cite{takahashi_2, takahashi_3, Santana2003b}.

In the nonrelativistic regime,  a general Galilean transformation acts on space and time as
\begin{equation}
    \mathbf{x}' = R\,\mathbf{x} + \mathbf{v}\,t + \mathbf{a}, 
    \qquad t' = t + \tau,
    \label{eq:gal}
\end{equation}
where $R\in SO(3)$, $\mathbf{v}$ is the inertial-frame velocity, and $\mathbf{a},\tau$ are shifts.
Velocities transform as $\dot{\mathbf{x}}' = R\,\dot{\mathbf{x}} + \mathbf{v}$ and
accelerations as $\ddot{\mathbf{x}}' = R\,\ddot{\mathbf{x}}$; hence acceleration is covariant
under rotations and insensitive to constant boost.

The Galilean symmetry can be represented by the generators satisfying
\begin{equation}
\begin{aligned}
    &[J_i,J_j]= i\hbar\,\varepsilon_{ijk}J_k,\quad
    [J_i,P_j]= i\hbar\,\varepsilon_{ijk}P_k,\quad
    [J_i,K_j]= i\hbar\,\varepsilon_{ijk}K_k,\\
    &[P_i,H]=0,\qquad [K_i,H]= i\hbar\,P_i, \qquad
    [K_i,P_j]= i\hbar\,m\,\delta_{ij},
\end{aligned}
\label{eq:bargmann}
\end{equation}
where $J_i$ stands for spatial rotations, $P_i$ represents spatial translations, $H$ the time translation, $K_i$ the \textit{boost}, and  $m$ is the \emph{central charge} (the total mass in the sector considered).
The presence of $m$ in $[K_i,P_j]$ characterizes the \emph{Bargmann central extension},
which is essential for a nontrivial quantum realization of the Galilei group.

In this framework, if we consider two vectors $\mathbf{x}=(x^1,x^2,x^3)$ and  $\mathbf{y}=(y^1,y^2,y^3)$ defined in the Euclidian space $\mathcal{E}$,  the quantity 
\begin{equation}\label{euc}
D^2= \mathbf{x}^2+\mathbf{y}^2-2\mathbf{x}\cdot\mathbf{y},
\end{equation}
is invariant under the transformations given in Eq.(\ref{eq:gal}) ($\mathbf{x}\cdot\mathbf{y}$ is the usual scalar product), and using Eq.(\ref{euc}), we can introduce a five-dimensional space in which the time can be included as a coordinate in the same level as spatial coordinates. This can be done using the concept of Galilean covariance. 

In this sense, Galilean covariance admits a geometric formulation within Newton–Cartan theory. Galilean covariance ensures that nonrelativistic laws retain their form under Eq.(\ref{eq:gal}). In order to introduce the Galilean covariance, we define the five-vectors 

\begin{equation}\label{fvec1}
x^\mu = (x^1,x^2,x^3,x^4,x^5) \equiv (\mathbf{x},t,s_x),
\end{equation}
and 
\begin{equation}\label{fvec1}
y^\mu = (y^1,y^2,y^3,y^4,y^5) \equiv (\mathbf{y},t,s_y)
\end{equation}
where 
\begin{equation}\label{c1}
x^4 = \frac{t - s_x}{\sqrt{2}}, \quad x^5 = \frac{t + s_x}{\sqrt{2}}, \qquad \textrm{and} \qquad y^4 = \frac{t - s_y}{\sqrt{2}}, \quad y^5 = \frac{t + s_y}{\sqrt{2}}.
\end{equation}

Vectors $x^{\mu}$ and $y^{\mu}$ are defined in Galilean space $\mathcal{G}$. In this way, the inner product in $\mathcal{G}$ is given by 
\begin{equation}
(x|y)=x^\mu y_\mu = \mathbf{x}\cdot\mathbf{y} - x^4 y^5 - x^5 y^4.
\end{equation}
Defining the metric tensor 
\begin{equation}
g_{\mu\nu} =
\begin{pmatrix}
\delta_{ij} & 0 & 0 \\
0 & 0 & -1 \\
0 & -1 & 0
\end{pmatrix},
\end{equation}
where $\delta_{ij}$ are the elements of the unitary matrix, it is possible to define the inner product as
\begin{equation}\label{inner}
(x|y)=g_{\mu\nu}x^\mu y^\nu.
\end{equation}
In the case of
\begin{equation}\label{ts2}
x^4=y^4=t, \qquad \textrm{and} \qquad x^5=\frac{\mathbf{x}^2}{2t}, \quad y^5=\frac{\mathbf{y}^2}{2t}.
\end{equation}
The inner product in $\mathcal{G}$ reduces to Eq.(\ref{euc}), i.e., $(x|y)=-\frac{1}{2}D^2$.
In this structure, the five-momentum is defined by
\begin{equation}
p^\mu = (-i\nabla,-i\partial_t,-i\partial_s) = (\mathbf{p},-E,-m).
\end{equation}
Then, the Galilei group in the covariant notation is defined by the transformations $\mathcal{G}: (x,t)\rightarrow (\overline{x},\overline{t})$ given by
\begin{eqnarray}\nonumber
\overline{x}^i&=&R^{i}_{j}x^{j}+v^{i}x^4+a^i,\\\nonumber
\overline{x}^4&=&x^4+a^4,\\\nonumber
\overline{x}^5&=&x^5+(R^{i}_{j}x^{j})v_i+\frac{1}{2}\mathbf{v}^2x^4,\nonumber
\end{eqnarray}
where $R^{i}_{j}$ stands rotations, $v^{i}$ stands boost, $a^i$ spatial translation, and $a^4$ time translation. The generators of this group can be written by $l_i=\frac{1}{2}\epsilon_{ijk}J_{jk}$, $k_i=M_{5i}$, $c_i=M_{4i}$, $d=M_{54}$, where $J_{ij}=x_ip_j-x_jp_i$. Using the operators given by $x_{\mu}=x_{\mu}\mathbf{1}$ and $p_{\mu}=-i\partial_{\mu}$ ($\hbar=1$), we can construct a representation for Galilei group in the context of Galilean covariance. For this purpose, we use the Casimirs of the Galilei algebra $I_1=p^{\mu}p_{\mu}$ and $I_2=p_5\mathbf{1}$. In this way, from invariant $I_2$ we obtain the equation

\begin{equation}
\partial_s \Psi = -\, i m\, \Psi
\end{equation}
 which the solution is given by
\begin{equation}
\Psi(\mathbf{x},t,s) = e^{-i m s}\,\psi(\mathbf{x},t).
\end{equation}
The invariant $I_1$ gives us the equation, 
\begin{equation}
\partial_\mu \partial^\mu \Psi = 0,
\end{equation}
which is a kind of Klein-Gordon equation for nonrelativistic particles.  In particular, the usual Schrödinger equation is obtained
\begin{equation}
i\,\partial_t \psi = -\,\frac{1}{2m}\,\nabla^2 \psi.
\end{equation}

In the same context, the lagrangian density that describes a nonrelativistic perfect fluid is given by 
\begin{equation}\label{lag1}
\tilde{\mathcal{L}}[\rho,\phi] = -\,\frac{1}{2}\,\rho\,\partial_\mu \phi\,\partial^\mu \phi \;-\; V(\rho).
\end{equation}
From Eq.(\ref{lag1}), Euler-Lagrange equations yields to 
\begin{equation}
\frac{1}{2}\,\partial_\mu \phi\,\partial^\mu \phi \;+\; V'(\rho) \;=\; 0,
\end{equation}
and 
\begin{equation}
\frac{1}{2}\,\nabla\phi\cdot\nabla\phi \;+\; \partial_t \phi \;=\; -\,V'(\rho).
\end{equation}
Also, we write the Euler (or continuity) equation
\begin{equation}
\partial_t \rho \;+\; \nabla\!\cdot(\rho\,\mathbf{v}) \;=\; 0.
\end{equation}
In addition, we get the Navier-Stokes equation
\begin{equation}
\partial_t \mathbf{v} \;+\; (\mathbf{v}\!\cdot\!\nabla)\,\mathbf{v} \;=\; -\,\frac{1}{\rho}\,\nabla p(\rho).
\end{equation}
In this way, the energy-momentum tensor
\begin{equation}
T^{\mu}{}_{\nu} =
\frac{\partial \mathcal{L}}{\partial(\partial_\mu \phi)}\,\partial_\nu \phi
- \delta^{\mu}{}_{\nu}\,\mathcal{L},
\end{equation}
satisfy the conservation law given by
\begin{equation}
\partial_\mu T^{\mu}{}_{\nu} = 0.
\end{equation}
Using Eq.(\ref{lag1}) we calculate the energy-momentum tensor for nonrelativistic fluid (notice that this tensor has, in general,  $25$-components)

\begin{align}
T_{\mu\nu}
&= 
\rho\!\left[
\frac{1}{2}(\partial_\alpha \phi)(\partial^\alpha \phi)\, g_{\mu\nu}
 - (\partial_\mu \phi)(\partial_\nu \phi)
\right]
 - V(\rho)\, g_{\mu\nu},
 \label{energia-momento}
\end{align}
where the field in the expression may be given by

\[
\phi(x,y,z, t, s) = \varphi(t) + \psi(x,y,z), \quad
\rho(x,y,z, t, s) = \varrho(t,s) \, \sigma(x,y,z)\,,
\]
which represents a separation of variables very useful for the content of the next section.

\section{Galilean Cosmology}\label{sec.3}

In the previous section, we reviewed the concept of Galilean covariance, namely how nonrelativistic fields can be described covariantly within a Galilean manifold. On the other hand, describing gravity in the same regime requires a generalization of this manifold: the Galilean manifold is then regarded as the flat background space. In this framework, nonrelativistic gravitation can be formulated using the same geometric tools of general relativity, but without the imposition of a universal speed limit on information propagation \cite{Ulhoa2009}. It is worth mentioning that the embedding procedure, characteristic of the Galilean manifold, is implemented by choosing a specific functional form for the coordinate s, which identifies the physical system under consideration. Once such a geometrical description of the gravitational interaction is established, it is natural to investigate cosmological solutions. Therefore, an isotropic and homogeneous universe becomes the immediate choice. In five dimensions, the line element with the desired symmetries, expressed in spherical coordinates, is given by
\begin{equation}
dS^2 = - (dx^4)^2 + a(x^4)^2 \Big(
\frac{dr^2}{1 - K r^2} + r^2 d\theta^2
+ r^2 \sin^2\theta\, d\phi^2 + r^2 \sin^2\theta \sin^2\phi\, d\gamma^2 \Big),
\end{equation}
where $K$ is the curvature and $a(x^4)$ is the scale factor. It is important to note that the isotropy and homogeneity of the metric were constructed from a five-dimensional Minkowski metric, which corresponds to the Galilean metric expressed in null coordinates, that is, by replacing $t$ and $s$ with $x^4$ and $x^5$. Therefore, the starting coordinates $x^\mu = (r,\theta,\phi,x^4,\gamma)$ must be transformed into $x'^\mu = (x,y,z,t,s)$. This is carried out in two intermediate steps. First, we consider the transformation
\begin{eqnarray}
r &=& \sqrt{x^2 + y^2 + z^2 + (x^5)^2},\nonumber\\
\theta &=& \arccos\!\left(\frac{z}{r}\right),\nonumber\\
\phi &=& \arctan\!\left(\frac{x}{\sqrt{y^2 + (x^5)^2}}\right),\nonumber\\
\gamma &=& \arctan\!\left(\frac{y}{x^5}\right)\nonumber
\end{eqnarray}

and then
\begin{eqnarray}
x^4 &=& \frac{t - s}{\sqrt{2}},\nonumber\\
x^5 &=& \frac{t + s}{\sqrt{2}}. \nonumber   
\end{eqnarray}
Once this structure is established, the Einstein tensor associated with the metric can be computed, allowing us to write down the field equations. In this formulation, the nonrelativistic fluid introduced in Sec.~\ref{sec.2} is physically interpreted in the coordinate system where $t$ indeed plays the role of the usual time variable. With these transformations, the components of the Einstein tensor can be written as

\begin{widetext}
\begin{align}
G_{11}
&=
\frac{\mathscr{E}}{K\,r^4(s+t)}\;
\Big[
- r^2\left(Kr^2 - 1\right)
\left(\tfrac{1}{2}(s+t)^2 + y^2 + z^2\right)
+ x^2\left(\tfrac{1}{2}(s+t)^2 + y^2 + z^2\right)
+ x^4
\Big]
\\[6pt]
G_{12}
&= x\,y\;\frac{2\,\mathscr{E}}{(s+t)}
\\[6pt]
G_{13}
&= x\,z\;\frac{2\,\mathscr{E}}{(s+t)}
\\[6pt]
G_{14}
&= G_{15}= x\,\mathscr{E}
\\[6pt]
G_{22}
&=
-\frac{2\,\mathscr{E}}{Kr^2(s+t)}\;
\Big[
(Kr^2-1)\Big(\tfrac12(s+t)^2+x^2+z^2\Big)
- y^2
\Big]
\\[6pt]
G_{23}
&= y\,z\,\frac{2\,\mathscr{E}}{(s+t)}
\\[6pt]
G_{24}
&= G_{25} = y\,\mathscr{E}
\\[6pt]
G_{33}
&=
-\frac{2\,\mathscr{E}}{Kr^2\,(s+t)}\,
\Big[
\big(Kr^{2}-1\big)\Big(\tfrac12(s+t)^{2}+x^{2}+y^{2}\Big)
- z^{2}
\Big]
\\[6pt]
G_{34}
&= G_{35}= z\,\mathscr{E}
\\[6pt]
G_{44}
&= \mathscr{H}
\;-\;
\frac{\mathscr{E}}{K\,r^4(s+t)}
\Big[
Kr^4(x^2+y^2+z^2)
- r^2(x^2+y^2+z^2)
- \tfrac{1}{2}(s+t)^2\, r^2
\Big]
\\[6pt]
G_{45}
&=
-\,\mathscr{H}
\;-\;
\frac{\mathscr{E}}{Kr^4\,(s+t)}
\Big[
K r^4(x^2+y^2+z^2)
- r^2(x^2+y^2+z^2)
- \tfrac{1}{2}(s+t)^2\,r^2
\Big]
\\[6pt]
G_{55}
&=
\mathscr{H}
-
\frac{\mathscr{E}}{Kr^4\,(s+t)}\;
\Big[
Kr^{4}(x^{2}+y^{2}+z^{2})
- r^{2}(x^{2}+y^{2}+z^{2})
- \tfrac12(s+t)^{2}\,r^{2}
\Big]\,,
\end{align}
\end{widetext}
where
\[
\mathscr{H}
:=
\frac{3}{A(s)^2\,\alpha(t)^2}\,
\left[
\tfrac12\!\left(A(s)\alpha'(t)-\alpha(t)A'(s)\right)^2
+K
\right]
\]
and
\begin{eqnarray}
\mathscr{E}
:=&&
\frac{3K(s+t)}{2(Kr^2 - 1)}
\Big[
\tfrac12\left(A(s)\alpha'(t)-\alpha(t)A'(s)\right)^2\nonumber\\
&&+\tfrac12 A(s)\alpha(t)\left(\alpha(t)A''(s)-2A'(s)\alpha'(t)+A(s)\alpha''(t)\right)
+K
\Big]\,.
\end{eqnarray}

As usual, the field equations are given by
\begin{align}
G_{\mu\nu} &= \kappa\,T_{\mu\nu},
\end{align}
where $T_{\mu\nu}$ is given by (\ref{energia-momento}), with the choice $V(\rho) = \lambda\rho$.
Since the off-diagonal components of the Einstein tensor are normally expressed in terms of $\mathscr{E}$, one sees that setting $K = 0$ implies that these components vanish. This leads to the conclusion that $\phi$ must be constant. In other words, this choice does not allow for an interacting fluid, since the gradient of the field $\psi$ determines the pressure of the fluid. On the other hand, experimental data indicate that the spatial curvature of the universe is very close to zero, and therefore there is an observational motivation for adopting $K = 0$. This does not exclude cosmological models with a pressured fluid, but it reduces the motivation for such a scenario given the current experimental constraints \cite{Planck2018}.

Therefore, we proceed by analyzing two different cases for the parameter $\lambda$, both corresponding to vanishing curvature. It should be mentioned that the parameter $\lambda$ controls the potential of the nonrelativistic fluid.

\subsection{Case 1: $\lambda=0$}

The first case corresponds to vacuum, since $\lambda = 0$ implies the vanishing of the energy--momentum tensor. Thus, the Einstein field equations reduce to

\begin{widetext}
\begin{align}
&4A'(s)\,\alpha'(t)
+
\frac{A(s)\!\left(\alpha(t)^{4}A'(s)^{2}-2\alpha'(t)^{2}\right)}{\alpha(t)}
+ A(s)^{2}\alpha(t)^{2}
\left(
-4A(s)A'(s)\alpha'(t)
+ \alpha(t)A''(s)
\right)
\nonumber\\[4pt]
&\quad
+ A(s)^{3}\alpha(t)
\left(
\alpha'(t)^{2}
+ \alpha(t)\alpha''(t)
\right)
=
\frac{2\alpha(t)A'(s)^{2}}{A(s)}\,,
\end{align}

\begin{align}
&\frac{2\alpha(t)A'(s)^{2}}{A(s)}
+
\frac{
A(s)\!\left(\alpha(t)^{4}A'(s)^{2}+2\alpha'(t)^{2}\right)
}{
\alpha(t)
}
+ A(s)^{2}\alpha(t)^{2}
\left(
-4A'(s)\alpha'(t)
+ \alpha(t)A''(s)
\right)
\nonumber\\[4pt]
&\quad
+ A(s)^{3}\alpha(t)
\left(
\alpha'(t)^{2}
+ \alpha(t)\alpha''(t)
\right)
=
4 A'(s)\,\alpha'(t)\,,
\end{align}

\begin{align}
A(s)^{2}\alpha'(t)^{2}
+ \alpha(t)^{2}
\big(
A'(s)^{2}
+ A(s)A''(s)
\big)
+ A(s)\alpha(t)
\big(
-4A'(s)\alpha'(t)
+ A(s)\alpha''(t)
\big)
= 0\,.
\end{align}
\end{widetext}

Combining these equations, we obtain the separable relation

\begin{equation}
\frac{\alpha''}{\alpha}
- \frac{{\alpha'}^{2}}{\alpha^{2}}
+ \frac{A''}{A} = 0\,.
\end{equation}
Since the functions depend independently on $t$ and $s$, we introduce a separation constant $\beta$, yielding

\begin{align}
\frac{A''}{A} &= \beta, \\[4pt]
\frac{\alpha''}{\alpha}
- \frac{{\alpha'}^{2}}{\alpha^{2}} &= -\beta.
\end{align}
The analytic solution is therefore

\begin{align}
A(s) &= C_{1}\,e^{\sqrt{\beta}\,s}
+ C_{2}\,e^{-\sqrt{\beta}\,s}, \\[4pt]
\alpha(t) &= C_{3}\,
\exp\!\left(
-\frac{\beta t^{2}}{2}
+ C_{4}t
\right).
\end{align}
This vacuum solution requires a physical interpretation. To that end, we reduce the five-dimensional space into the usual 3+1 decomposition by choosing
\[
s = \frac{\bar{r}^{2}}{2t}\,,
\qquad
\text{where } \bar{r}^{2} = x^{2}+y^{2}+z^{2}.
\]
It is well known that such a choice for the coordinate $s$ embeds the Euclidean space into the five-dimensional Galilean manifold. We emphasize that the dimensional adjustment constant $\bar{c}$ is set equal to 1, corresponding to a natural system of units appropriate to this manifold.

With this prescription, the scale factor $a = A\,\alpha$ becomes dependent on both time and the spatial coordinates. This naturally introduces anisotropy into the model due to the dimensional reduction. In other words, the isotropy and homogeneity of the five-dimensional space are broken when the theory is projected down to the nonrelativistic 3+1 description.

Such anisotropy is consistent with the observed features of our universe and may be tuned through suitable choices of the integration constants in our approach.

\subsection{Case 2: $\lambda=1$}

We now consider the second case, in which the parameter $\lambda$ assumes a finite value. As discussed previously, $\lambda$ controls the potential associated with the nonrelativistic fluid and, therefore, this case describes a universe filled with matter, defined by the density $\rho$. In this framework, the field equations take the following form

\begin{widetext}
\begin{align}
\frac{3}{4} \Bigg(
&\frac{2 A'(s)^2}{A(s)^2}
- \alpha(t)^2 A'(s)^2
- \frac{4 A'(s)\,\alpha'(t)}{A(s)\,\alpha(t)}
+ \frac{2 \alpha'(t)^2}{\alpha(t)^2}
\Bigg)
+ \frac{3}{4} A(s)\,\alpha(t)
\left(
4 A'(s)\,\alpha'(t)
- \alpha(t)\,A''(s)
\right)
\nonumber
\\
&
- \frac{3}{4} A(s)^2
\left(
\alpha'(t)^2
+ \alpha(t)\,\alpha''(t)
\right)
=
-\frac{1}{2}
\left(
-1
+ A(s)^2\,\alpha(t)^2
\right)
\varrho(t,s)\,
\kappa \sigma(x,y,z)\,,
\end{align}

\begin{align}
\frac{3}{4}\Bigg(
&-\frac{2\,A'(s)^2}{A(s)^2}
+ \frac{4\,A'(s)\,\alpha'(t)}{A(s)\,\alpha(t)}- \frac{\alpha(t)^4\,A'(s)^2 + 2\,\alpha'(t)^2}{\alpha(t)^2}
\Bigg) + \frac{3}{4} A(s)\,\alpha(t)\!\left(4 A'(s)\,\alpha'(t) - \alpha(t)\,A''(s)\right)
\nonumber\\
&
- \frac{3}{4} A(s)^2\!\left(\alpha'(t)^2 + \alpha(t)\,\alpha''(t)\right)
=
-\frac{1}{2}\,\big(1 + A(s)^2\,\alpha(t)^2\big)\,
\varrho(t,s)\, \kappa \sigma(x,y,z)
\end{align}
and
\begin{align}
&A(s)^2 \Big(
2\,\alpha(t)^2\,\varrho(t,s)\, \kappa \sigma(x,y,z)
- 3\,\alpha'(t)^2
- 3\,\alpha(t)\,\alpha''(t)
\Big)\nonumber\\
&=
3\,\alpha(t)\Big(
-4\,A(s)\,A'(s)\,\alpha'(t)
+ \alpha(t)\,\big(A'(s)^2 + A(s)\,A''(s)\big)
\Big)\,.
\end{align}
\end{widetext}
Thus, by means the choice of $\kappa \sigma(x,y,z) =1$ and after some algebraic manipulations, we arrive at

\begin{equation}
\frac{\alpha''}{\alpha} - \frac{\alpha'^2}{\alpha^2}
+ \frac{A''}{A} - \frac{A'^2}{A^2} = 0\,,
\label{eq:base}
\end{equation}
which could be expressed as

\begin{equation}
\frac{d}{ds}\left(\frac{A'}{A}\right) = \beta
\qquad\text{and}\qquad
\frac{d}{dt}\left(\frac{\alpha'}{\alpha}\right) = -\beta
\label{eq:derivatives}
\end{equation}
whose solutions are
\begin{equation}
A(s) = A_0\, e^{\beta s^2 + A_1 s}
\label{eq:Asol}
\end{equation}
and
\begin{equation}
\alpha(t) = \alpha_0\, e^{-\beta t^2 + \alpha_1 t}\,.
\label{eq:alphasol}
\end{equation}
Therefore the fluid density reads
\begin{equation}
\varrho(s,t)=
3\left(
2\beta(s+t)+A_1-\alpha_1
\right)^2\,.
\label{rho_final}
\end{equation}
Similarly to Case 1, the choice $s = \frac{\bar{r}^{2}}{2t}$ establishes both the scale factor $a$ and the fluid density as anisotropic quantities. If the curvature $K$ were different from zero, anisotropy would also arise from the fluid pressure. In this scenario, the fluid density becomes the directly relevant experimental quantity, since its measurement allows one to infer the physical properties of the fluid. Observational data then suggest that matter is distributed throughout the universe without interaction.

\section{Conclusion} \label{sec.4}

We have developed a nonrelativistic cosmological model grounded in Galilean covariance and formulated within a five-dimensional manifold. The construction of an isotropic and homogeneous metric analogous to the Friedmann--Robertson--Walker line element allowed us to derive Einstein-like field equations without assuming a relativistic speed limit. Two physically distinct solutions were obtained. The first one, corresponding to the vacuum case (\(\lambda=0\)), yields an exponential--quadratic dependence for the scale factor. The second solution (\(\lambda=1\)) represents a dust-filled universe described by a non-interacting nonrelativistic fluid. Through the embedding procedure that reduces the manifold to 3+1 spacetime, the model naturally acquires anisotropy in both the scale factor and the density, while remaining consistent with the near-flat spatial curvature indicated by cosmological observations. In the particular case of vanishing curvature, the model reproduces Milne’s Newtonian cosmology because this condition leads to a vanishing pressure. In Newtonian cosmology, since Milne’s seminal formulation, the dynamics is regarded as an approximation to General Relativity. The system is described as matter expanding within a fixed Euclidean space, in contrast to the relativistic picture of an expanding spacetime. This simplification, while conceptually appealing, was restricted to the case of zero curvature and vanishing pressure, thus producing a strictly isotropic model. Subsequent neo-Newtonian extensions introduced pressure, yet maintained isotropy as a built-in assumption. In the Galilean framework developed here, isotropy and homogeneity are initially imposed in the five-dimensional metric, but the dimensional reduction to 3+1 spacetime naturally breaks isotropy in the Galilean coordinates. As a result, the model exhibits anisotropy of geometric origin, without invoking pressure or additional dynamical sources. This feature provides a more realistic description of the universe, since small anisotropies are indeed observed, while remaining compatible with a nearly flat spatial curvature. Therefore, the Galilean cosmology presented here extends Newtonian formulations beyond the limits of the relativistic approximation, offering an independent and covariant geometric framework for nonrelativistic gravitation. It preserves the simplicity of Newtonian cosmology while naturally incorporating anisotropic effects through its geometric structure. Future work will address extensions of this model to include curvature-dependent pressure and the thermodynamic aspects of the Galilean gravitational field.

\section*{Acknowledgments}

This work by A. F. S. is partially supported by National Council for Scientific and Technological
Development - CNPq project No. 312406/2023-1.

\section*{Data Availability Statement}

No Data associated in the manuscript.


\end{document}